\documentclass[10pt]{article}    

\usepackage{amssymb,latexsym,amsmath} 
\usepackage[mathscr]{eucal}  
\newfont{\sffl}{msbm10 at 16pt} 
\newfont{\sff}{msbm10 at 10pt}

\begin{document}           
\title{\vskip -.75in
Comments on the Reliability of Lawson and Hanson's\\
    Linear Distance Programming Algorithm:\\
 Subroutine LDP
\thanks{\small{Approved for public  release.}}}

\author{Alan Rufty\\         
\\
Naval Surface Warfare Center Dahlgren\\
Attn: Alan Rufty\\
 5443 Bronson Road, Suite 112\\
Dahlgren, VA. 22448--5159}
\maketitle                 

\newcommand{\KD}{K_{\text{D}}}

\begin{abstract}
  This brief paper: (1) Discusses strategies to generate random test cases that can be used to extensively test any Linear Distance Program (LDP) software.  (2) Gives three numerical examples of input cases generated by this strategy that cause problems in the Lawson and Hanson LDP module.  (3) Proposes, as a standard matter of acceptable implementation procedures, that (unless it is done internally in the software itself, but, in general, this seems to be much rarer than one would expect) all users should test the returned output from any LDP module for self-consistency since it incurs only a small amount of added computational overhead and it is not hard to do.

\vskip .1in
\noindent
\textbf{ACM} Categories and Subject Descriptors:  G.4 [\textbf{Mathematical Software}]: Reliability and robustness; D.2.5 [\textbf{Testing and Debugging}]: Testing tools (e.g., data generators, coverage testing)
\end{abstract}

\newcommand{\SubSec}[1]{

\vskip .18in
\noindent
\underline{{#1}}
\vskip .08in

}
\newcommand{\ls}{\vphantom{\big)}} 
\newcommand{\lsm}{\!\vphantom{\big)}} 

\newcommand{\eq}{:=}

\newcommand{\smallindex}[1]{\text{\raisebox {1.5pt} {${}_{#1}$}}}

\newcommand{\mLarge}[1]{\text{\begin{Large} $#1$ \end{Large}}}
\newcommand{\mSmall}[1]{\text{\begin{footnotesize} $#1$ \end{footnotesize}}}

\newcommand{\Blbrac}{\rlap{\bigg{\lceil}}\bigg{\lfloor}} 
\newcommand{\Brbrac}{\bigg{\rbrack}} 

\newcommand{\Dr}{\mathscr{D}_r}

\newcommand{\smallindexes}[1]{\text{\raisebox {.5pt} {${}_{#1}$}}}

\newcommand{\hc}{\text{H}_{{\text{\raisebox {1.5pt} {${}_{\mspace{-1.3mu}\perp}$}}}}^{\mspace{1.5mu}\text{\raisebox {1pt} {$c$}}}}
\newcommand{\Hc}{\text{H}^c}

\vskip .001in
\noindent
\begin{itemize}
\item[\ \ ] \small{\textbf{Key words:} {Linear Distance Programming, Matrix Inequality, Kuhn-Tucker Theorem}}
\item[\ \ ] \small{\textbf{AMS subject classification (2000):} {Primary 15A39. Secondary 15A45}}
\end{itemize}

\renewcommand {\baselinestretch}{1.35} 
 
\section{Introduction}\label{S:intro}

Given some $m \times n$ matrix $\mathbf{G}$ and $m$-vector $\mathbf{h}$, the Linear Distance Programming (LDP) problem consists of finding the $n$-vector $\mathbf{X}$ of minimum norm that satisfies the following system of inequality constraints (when a solution exists):
\begin{equation}\label{E:ldp}
 \mathbf{G}\mathbf{X} \geq \mathbf{h}\ .
\end{equation}
Here the phrase ``$\mathbf{X}$ is of minimum norm'' means that $|\mathbf{X}|$ is to be minimized, subject to the set of constraints given by (\ref{E:ldp}) and provided that these constraints are consistent.

  A properly implemented LDP algorithm must thus:
\begin{enumerate}
 \item
 find the $\mathbf{X}$ of minimum norm, when the system of constrains given by (\ref{E:ldp}) is consistent or
\item
 return a flag indicating that the set of constraints given by (\ref{E:ldp}) is inconsistent and thus that no solution exists.
\end{enumerate}
The very well known and extensively utilized Lawson and Hanson software module ``Subroutine LDP'' is an implementation of this LDP problem \cite{LandH1,LandH2}.   The distributions of the Lawson and Hanson algorithm(s) are generally in Fortran and can be ported from the web site netlib at:

\noindent
\ \ \ \ http://netlib.org/lawson-hanson/all\ .

\noindent
These distributions are also available at numerous other web sites such as
 
\noindent
\ \ \ \ http://netlib.bell-labs.com/lawson-hanson/index.html\ .

\noindent

 When the inequality constraints are determined to be consistent, the Lawson and Hanson module LDP sets the flag MODE to 1 and returns a solution
vector $\mathbf{X}$.  Alternatively, if the constraints are determined to be inconsistent, the Lawson and Hanson module LDP sets the flag MODE to 4 indicating that there is no solution.  The prime concern here is the relatively rare occurrence of a MODE = 1 return with an associated solution vector $\mathbf{X}$ that fails to satisfy the set of constraints (\ref{E:ldp}).  Although these problem cases seem to be associated with issues in handling significant digits, the erroneous returned solutions may, in fact, produce numerically sizeable violations of the specified inequality constraints.

  These types of problem examples were uncovered using a strategy of systematically generating copious numbers of random cases.  This general strategy is outlined in Section~2 and three specific numerical examples generated by this strategy that cause problems and have small values of $m$ and $n$ are given in Section~3 and then discussed in Section~4.

 Although the Lawson and Hanson software LDP module \cite{LandH1,LandH2} is of primary interest here, the general checkout strategies and recommendations apply to other LDP implementations as well.   In particular similar testing was performed on the NSWC Mathematics Library Subroutine LSEI \cite{NSWCML}, which can be used to handle LDP problems.  This testing revealed that LSEI was also prone to yield a MODE = 1 return under rare circumstances when the set of constraints (\ref{E:ldp}) was inconsistent; however, the properties of the $\mathbf{G}$ matrices that caused problems appeared to be different for LSEI and LDP.  In this regard, the remark at the bottom of page 393 of \cite{NSWCML} may be relevant and is worth repeating here:
\begin{enumerate}
\item
``LSEI may perform poorly if the norms of the rows of $\mathbf{A}$ and $\mathbf{E}$ differ by many orders of magnitude, or if the norms of the rows of $\mathbf{E}$ are exceedingly small.'' 
\end{enumerate}

  Given that problems were found for the only two LDP algorithms tested and because it is clearly easy to do and entails only a small added computational burden, as a 
matter of course, it is recommended that users of any LDP software module should automatically check the constraint condition (\ref{E:ldp}) of all returned solutions when the software module indicates that there is one.
 
\section{Random Case Generation Strategy}\label{S:random}

  One significant point is that random cases that are guaranteed to satisfy some inequality constraint or other can be generated systematically.  First, consider the set of inequality constraints that results for the choice $\mathbf{h} = 0$:
\begin{equation}\label{E:ldp2}
 \mathbf{G}\mathbf{X} \geq 0\ .
\end{equation}
Clearly, in this case, $\mathbf{X} = 0$ is always the solution to the LDP problem since $|\mathbf{X}| = 0$.

  It is easy to generalize this solution.  Towards that end consider the set of conditions
\begin{equation}\label{E:ldp3}
 \mathbf{G}(\mathbf{X} - \mathbf{X}_0) \geq 0\ ,
\end{equation}
where $\mathbf{X}_0$ is some (randomly) chosen vector.  Clearly (\ref{E:ldp3}) always specifies a set of consistent constraints since $\mathbf{X} = \mathbf{X}_0$ is always a valid solution to the constraint conditions.  Some other choice of $\mathbf{X}$ with  $|\mathbf{X}| < |\mathbf{X}_0|$ may exist so $\mathbf{X} = \mathbf{X}_0$ may not be the actual solution to the LDP problem, but at least a solution is guaranteed.  Thus, for any randomly chosen vales of $m$ and $n$, if $\mathbf{G}$ is a randomly generated $m \times n$ matrix and $\mathbf{X}_0$ is a randomly generated $n$-vector, then provided the $n$-component $\mathbf{h}$ vector is computed by
\begin{equation}\label{E:ldp4}
 \mathbf{h} := \mathbf{G}\mathbf{X}_0\ ,
\end{equation}
a consistent solution to (\ref{E:ldp}) always exists.

  This whole procedure can be carried one step further to obtain a more general set of self-consistent equations of the form (\ref{E:ldp}).  Let $\mathbf{A}$ be a general (i.e., randomly generated) $l \times m$ matrix and $\mathbf{B}$ be an arbritrary invertible (i.e., also possibly randomly generated) $n \times n$ matrix.  Then (\ref{E:ldp3}) yields the following form
\begin{equation}\label{E:ldp5}
 \mathbf{A}\mathbf{G}\mathbf{B}\mathbf{B}^{-1}(\mathbf{X} - \mathbf{X}_0) \geq 0\ ,
\end{equation}
which can be rewritten as 
\begin{equation}\label{E:ldp6}
 \mathbf{G}'\mathbf{X}' \geq \mathbf{h}'
\end{equation}
where 
\begin{equation}\label{E:ldp7}
 \mathbf{G}' := \mathbf{A}\mathbf{G}\mathbf{B}\ \ \text{and}\ \ \ \mathbf{h}'
:= \mathbf{A}\mathbf{G}\mathbf{X}_0\ .
\end{equation}
Here the solution $\mathbf{X}'$ is related to $\mathbf{X}$ by $\mathbf{X}' = \mathbf{B}^{-1}\mathbf{X}$ and the inclusion of $\mathbf{B}$ here allows for additional freedom in the choice of $\mathbf{X}$ since it helps to de-link the solution vector from $\mathbf{X}_0$.

  By generating a random $l, m$, and $n$ and then appropriate random matrices $\mathbf{A}$, $\mathbf{B}$ and $\mathbf{G}$, along with a random vector $\mathbf{X}_0$, it is easy to test out any LDP algorithm for the case of self-consistent constraints using (\ref{E:ldp7}).  Notice that in practice it is easy to generate a non-singular $\mathbf{B}$ matrix with random elements:  Simply generate completely random $\mathbf{B}$ matrices and then test the result to see if $|\mathbf{B}| = 0$ and then simply regenerate the $\mathbf{B}$ matrix if it is.  (Appropriate scaling factors, as well as other minor implementation issues are obvious from the examples in Section~\ref{S:random}.)
    
  Observe that since the existence of a solution in the above derivation was predicated on the existence of equalities [i.e., the $\geq$ and not $=$ in, say (\ref{E:ldp}) or (\ref{E:ldp7})], the region where the constraints are consistent may consist of only a single point and thus (due to round-off or other factors) the LDP algorithms may potentially experience some problems in finding and testing for this one point.  One way to enlarge this interior feasibility region is to replace (\ref{E:ldp}) by
\begin{equation}\label{E:ldp8}
 \mathbf{G}\mathbf{X} \geq \tilde{\mathbf{h}}\ ,
\end{equation}
where $\tilde{\mathbf{h}} = {\mathbf{h}} - \mathbf{C}$, for some constant or random vector $\mathbf{C}$, all of whose components are positive.  It is also easy to modify the above random case generation strategy so that inconsistent constraints are produced.  One strategy, for example, is to simply replace $\tilde{\mathbf{h}}$ in (\ref{E:ldp8}) by $\tilde{\mathbf{h}} = {\mathbf{h}} + \mathbf{D}$ for some random vector $\mathbf{D}$ with sufficiently large components.
 
\section{Numerical Examples of Randomly Generated Problem Cases}\label{S:examples}
    
  Three numerical examples of inconsistent inequality constraints are given below.  When the input for each of these cases is used in the Lawson and Hanson Fortran Subroutine LDP implemented on a SUN workstation using \emph{double precision}, a MODE = 1 return results and a candidate solution vector is returned, whereas there should be a MODE = 4 return indicating that no solution exists.  Specific problem cases are by nature somewhat finicky and difficult to replicate and may not cause similar problems on other computer systems---they are, in fact, highly dependent on the number of internal significant digits used for the inputs, as well as in the internal computations themselves.  (SUN architecture reuses the mantissa when going from single to \emph{double precision} so there are generally 16 or so significant digits of internal representation rather than the 14 or so one might normally expect.)  This, however, does not mean that the problem is tied to only SUN systems or that they are more common on SUN systems; rather, it means that different specific cases will cause problems on different systems. Given this state of affairs, the reader may well not be able to use the input cases below to generate inconsistent returns from Subroutine LDP and may thus have to use the strategies outlined above to generate such cases randomly.  [Generally, $\mathbf{A}$ and $\mathbf{B}$ can be taken to be identity matrices, but a certain number of columns of the random $\mathbf{G}$ should be zero and since (for what might be typical reasonable parameter settings designed to generate such cases and an associated well-chosen random test case generation methodology) the incidences of such problem cases might usually be only approximately one in a thousand or so, probably at least 50,000 or so cases of various types and with parameter settings should normally be generated.]  In passing, it is worth noting that the problems addressed here are not tied specifically to Fortran: a C programming language version of the Lawson and Hanson LDP algorithm was also tested and displayed the same problems.

  Finally, before considering the examples, two observations are in order.  First, notice that the Lawson and Hanson in-line function ``DIFF'' is used in their LDP implementation to help prevent internal significant digit masking, but this, in itself, does not appear to resolve the problems indicated here.  Second, when these same inputs are used with a \emph{type quad} internal Fortran version (which corresponds to approximately 32 internal digits of representation), then they all yield MODE = 4 returns, but it seems clear that simply going to a \emph{type quad} implementation will not solve the problem (although it will lessen the incidence of it)---it will only make counter-examples harder to find.

  Although the cases $m >> n >> 0$ are the ones that generate the most problem cases, for simplicity only three cases with $m = 4$ and $n = 2$ are considered.

\vskip .2in
\noindent
Case 1
\begin{equation}\label{E:realT}
\mathbf{G} =
\left[
\begin{array}{r l}
 -89.20509815216064 & 0.0000000000000000\\ 
  74.79768991470337 & 0.0000000000000000\\ 
  66.23740792274475 & 0.0000000000000000\\ 
 -18.51919293403625 & 0.0000000000000000\\ 
\end{array}  
\right]\! \ \ \ \ \ \ 
\mathbf{h} =
\left[
\begin{array}{r}
 -12073.43407295207\\
  10123.19482867013\\
  8350.549301112449\\
 -24612.94532321187\\
\end{array}
\right]
\end{equation}

\vskip .2in
\noindent
Case 2
\begin{equation}\label{E:realT2}
\mathbf{G} =
\left[
\begin{array}{r l}
 81.82253837585449 & 0.0000000000000000\\ 
 -74.02672171592712 & 0.0000000000000000\\ 
  0.0000000000000000 & -17.36225485801697\\ 
 -89.47155475616455 & 0.0000000000000000\\ 
\end{array}  
\right]\! \ \ \ \ \ \ 
\mathbf{h} =
\left[
\begin{array}{r}
 -77004.09890544150\\
  69248.37468031116\\
 11241.52852765946\\
 84233.37742495652\\
\end{array}
\right]
\end{equation}

\vskip .2in
\noindent
Case 3
\begin{equation}\label{E:realT3}
\mathbf{G} =
\left[
\begin{array}{r l}
 -3.057897090911865 & 0.0000000000000000\\ 
  4.310655593872070 & 0.0000000000000000\\ 
  39.13614749908447 & 0.0000000000000000\\ 
  84.55699086189270 & 0.0000000000000000\\ 
\end{array}  
\right]\! \ \ \ \ \ \ 
\mathbf{h} =
\left[
\begin{array}{r}
 2192.778913749731\\
 -3422.354440768027\\
 -28760.98260603488\\
 -60562.89687439907\\
\end{array}
\right]
\end{equation}
 
\section{Discussion of Randomly Generated Problem Case Outputs}\label{S:random2}
 
  For each of the above examples, it is easy to verify directly that the system of constraints implied by (\ref{E:ldp}) is consistent; however, when the input for each of these examples was used  by the author in the Lawson and Hanson Subroutine LDP implemented in \emph{double precision} on a SUN system, a MODE = 1 return resulted with an improper solution vector.  Specifically the returned solution vectors were:

\vskip .2in
\noindent
Case 1
\begin{equation}\label{E:real2} 
\mathbf{X} =
\left[
\begin{array}{r}
  -.3750000000000000\\
  .0000000000000000\\
\end{array}
\!\!\!\!\!\!
\begin{array}{c}
\\ 
\\
\\
\end{array}
\right]
\end{equation}

\vskip .2in
\noindent
Case 2
\begin{equation}\label{E:realT5} 
\mathbf{X} =
\left[
\begin{array}{c}
  .6074218750000000\\
  .0000000000000000\\
\end{array}
\!\!\!\!
\begin{array}{c}
\\ 
\\
\\
\end{array}
\right]
\end{equation}

\vskip .2in
\noindent
Case 2
\begin{equation}\label{E:realT6} 
\mathbf{X} =
\left[
\begin{array}{c}
  .4570312500000000\\
  .0000000000000000\\
\end{array}
\!\!\!\!
\begin{array}{c}
\\ 
\\
\\
\end{array}
\right]
\end{equation}

 For these cases, direct numerical testing shows that the problems seem to be in the Subroutine LDP itself and not in the other subroutines it calls.  More specifically, using Lawson and Hanson's notation \cite{LandH1,LandH2}, except for the use of bold face type for vectors and matrices, from the returns supplied to Subroutine LDP from Subroutine NNLS (which LDP calls), the basic vectors and matrices $\mathbf{E}$, $\mathbf{r}$, $\mathbf{p}$ and $\hat{\mathbf{x}}$ can be determined.  \{Specifically, see equations (23.28) through (23.34) in \cite{LandH1} or \cite{LandH2}.\}  Using these computed quantities it easy to show that due to round-off and/or masking the basic assumptions in proving the validity of the LDP algorithm do not hold; For example, that some of the components of the $\mathbf{p}$ vector are negative.  For these specific inputs all of these inconsistencies go away when \emph{type quad} variables are used, but as indicated above, one would expect to be able to find similar cases when a full set of \emph{type quad} random cases (one tricky part here appears to be finding a full-up \emph{type quad} random number generator).  Finally, with regards to  \emph{type quad} Fortran implementations, it is perhaps worth noting that there is a SUN $f77$ compiler option that automatically converts from \emph{type double} to \emph{type quad}.


\end{document}